# FAST MULTI-ENCODING TO REDUCE THE COST OF VIDEO STREAMING

*Hadi Amirpour[1], Vignesh V Menon[1], Ekrem Çetinkaya[1], Adithyan Ilangovan[2], Christian Feldmann[2], Martin Smole[2], and Christian Timmerer[1,2]*

[1]Christian Doppler Laboratory ATHENA, Alpen-Adria-Universität, Klagenfurt, Austria
{*firstname.lastname*}@aau.at

[2]Bitmovin, Klagenfurt, Austria
{*firstname.lastname*}@bitmovin.com

## ABSTRACT

The growth in video Internet traffic and advancements in video attributes such as framerate, resolution, and bit-depth boost the demand to devise a large-scale, highly efficient video encoding environment. This is even more essential for Dynamic Adaptive Streaming over HTTP (DASH)-based content provisioning as it requires encoding numerous representations of the same video content. *High Efficiency Video Coding* (HEVC) is one standard video codec that significantly improves encoding efficiency over its predecessor *Advanced Video Coding* (AVC). This improvement is achieved at the expense of significantly increased time complexity, which is a challenge for content and service providers. As various representations are the same video content encoded at different bitrates or resolutions, the encoding analysis information from the already encoded representations can be shared to accelerate the encoding of other representations. Several state-of-the-art schemes first encode a single representation, called a reference representation. During this encoding, the encoder creates analysis metadata with information such as the slice-type decisions, CU, PU, TU partitioning, and the HEVC bitstream itself. The remaining representations, called dependent representations, analyze the above metadata and then reuse it to skip searching some partitioning, thus, reducing the computational complexity. With the emergence of cloud-based encoding services, video encoding is accelerated by utilizing an increased number of resources, *i.e.*, with multi-core CPUs, multiple representations can be encoded in parallel. This paper presents an overview of a wide range of multi-encoding schemes with and without the support of machine learning approaches integrated into the HEVC Test Model (HM) and x265, respectively. Seven multi-encoding schemes are presented, and their performance in encoding time complexity and bitrate overhead compared to the state-of-the-art approaches are shown. Enabling fast multi-encoding for HAS in modern *Over-the-top* (OTT) workflows will reduce time-to-market and costs immensely.

## INTRODUCTION

*HTTP Adaptive Streaming* (HAS) is the de-facto standard in delivering videos over the internet to a variety of devices. The main idea behind HAS is to divide the video content into segments and encode each segment at various bitrates and resolutions, called representations, which are stored in plain HTTP servers as shown in Figure 1. These representations are stored in order to continuously adapt the video delivery to the network conditions and device capabilities of the client. To meet the high demand for streaming high-quality video content over the Internet and overcome the associated challenges in HAS, the *Moving Picture Experts Group* (MPEG) has developed a standard called *Dynamic Adaptive Streaming over HTTP* (MPEG-DASH) [1]. The increase in video traffic and improvements in video characteristics such as resolution, framerate, and bit-depth raise the need to develop a large-scale, highly efficient video encoding environment [2]. This is even more crucial for DASH-based content provisioning as it requires encoding multiple representations of the same video content [3].

*High Efficiency Video Coding* (HEVC) [4] is one standard video codec that is widely being used in content production nowadays. Based on Bitmovin's video developer report in 2021 [5], HEVC is used in 49% of productions in 2021 and it is expected to be added to more than 25% of extra productions in 2022. HEVC significantly improves coding efficiency over its predecessor *Advanced Video Coding* (AVC) [6]. This improvement is achieved at the cost of significantly increased runtime complexity, which is a challenge for content and service providers. As various representations of the same video content are encoded at different bitrates or resolutions, the encoding analysis information from the already encoded representations can be shared to accelerate the encoding of other representations.

Several state-of-the-art schemes [6–12], first encode a single representation, called a reference representation. The encoder creates analysis metadata (file) with information such as the slice-type decisions [13], CU, PU, TU partitioning [14], and the HEVC bitstream itself during this encoding. The remaining representations, called dependent representations, analyze the above metadata and then reuse it to skip searching some partitioning, thus, reducing the computational complexity. With the emergence of cloud-based encoding services [15] and in live applications video encoding is accelerated by utilizing an increased number of resources, *i.e.*, with multi-core CPUs, multiple representations can be encoded in parallel.

In this paper, the schemes are analyzed for both serial and parallel encoding environments. The term multi-rate is used when all representations are encoded at a single resolution but at different bitrates. Multi-encoding is used when a single video is provided at various resolutions, and each resolution is encoded at different bitrates.

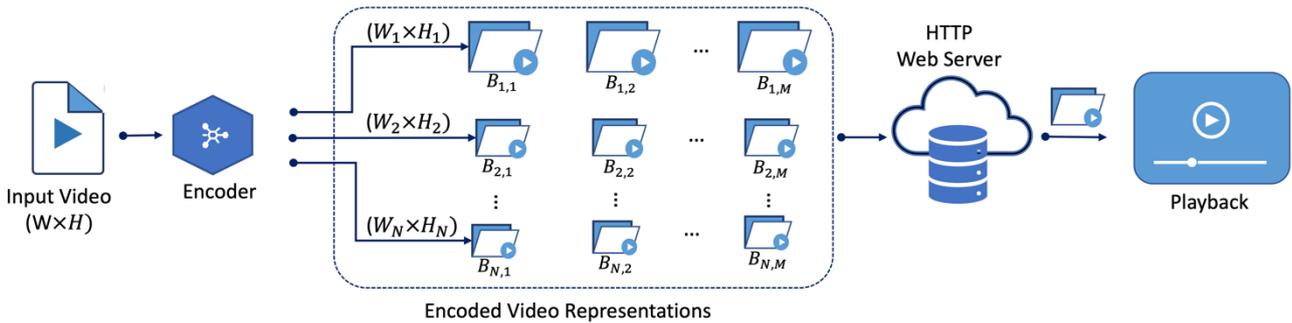

Figure 1. Encoded video representations as used in DASH. $W$ and $H$ denote the frame width and frame height, respectively. $N$ denotes the number of resolutions stored. $W_1 \times H_1$ and $W_N \times H_N$ denote the lowest and highest resolutions, respectively. $M$ denotes the number of bitrate representations in each resolution. $B_{i,1}$ to $B_{i,M}$ represent the target bitrates in the ascending order for the representations in each resolution.

## BACKGROUND AND RELATED WORK

In HAS, the same video content is encoded at multiple bitrates to continuously adapt the video delivery to the users' need, resulting in a significant increase in the encoding cost. However, as the same content is encoded multiple times, the encoder analysis information from the already encoded representation(s) can be reused to speed up the encoding process of the remaining representations. In HEVC, frames are first divided into slices, and then they are further divided into square regions called Coding Tree Units (CTU) [4], which are the main building blocks of HEVC. To encode each CTU, it is recursively divided into smaller square regions called Coding Units (CUs) (see Figure 2). Depth values from 0 to 3 are assigned to CU sizes from $64 \times 64$ to $8 \times 8$ pixels. Therefore, to find the best CTU partitioning, 85 CUs including one $64 \times 64$ CU, four $32 \times 32$ CUs, sixteen $16 \times 16$ CUs, and sixty-four $8 \times 8$ CUs are searched.

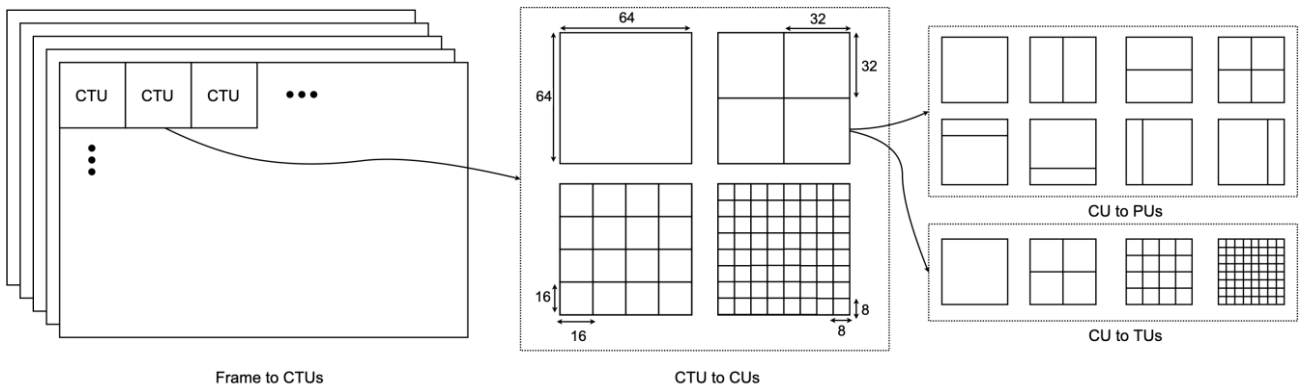

Figure 2. In HEVC, frames are divided into CTUs, and each CTU is then divided into CUs. Each CU is subdivided into PUs, and the prediction residuals of CUs are partitioned into TUs. The optimal CTU partitioning is found after an exhaustive search process through all CUs, PUs, and TUs.

The inter-prediction modes comprise of Merge/Skip 2N×2N, Inter 2N×2N, Symmetric Motion Partition (SMP, including Inter 2N×N and Inter N×2N), Asymmetric Motion Partition (AMP, including Inter 2N×nU, Inter 2N×nD, Inter nL×2N, and Inter nR×2N), and Inter N×N. In contrast, the intra-prediction modes involve Intra 2N×2N and Intra N×N. The best PU mode is selected according to all modes' minimum rate-distortion cost (RD-cost). Furthermore, for transform coding of the prediction residuals, each CU can be partitioned into multiple transform blocks (TBs), the size of which can also vary from 4x4 to 32x32 pixels [4]. In general, finding the optimal partitioning for each CTU is time-consuming, given the possible CU, PU, and TU partitioning, multiple reference frame search processes for inter-coded CUs, and the actual motion estimation algorithm.

**MULTI-RATE (MR) SCHEMES**

Schroeder et al. [7] propose a single bound approach for CU depth estimation, in which, firstly, the highest bitrate representation is encoded (see Figure 3). Its CU depth information is then used to encode the other representations. CUs generally have higher depth values in higher bitrate representations. The CU depth search range for the representations, $B_{i,j}$ $\forall j \in [1, M]$ of the $(i)^{th}$ resolution is calculated as:

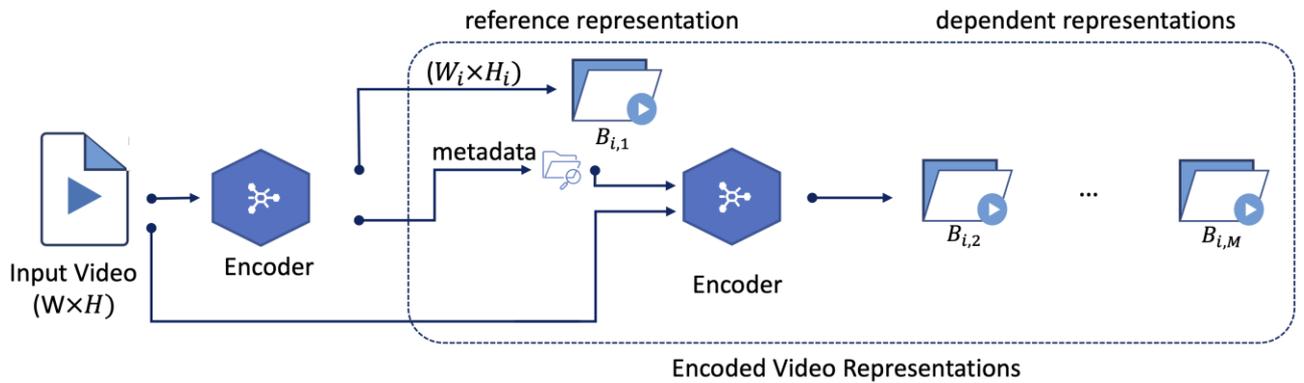

Figure 3. Encoder analysis flow diagram of [7] for M representations. The target bitrates for each representation, $B_{i,1}$ to $B_{i,M}$ are arranged in ascending order.

$$[d_L, d_U] = \begin{cases} [0, 3], & B_{i,M} \\ [0, d_{i,M}], & B_{i,j} \forall j \in [1, M-1] \end{cases} \quad (1)$$

where $M$ represents the number of bitrate representations in the $(i)^{th}$ resolution, $B_{i,M}$ is the highest bitrate representation of the $(i)^{th}$ resolution and $d_{i,M}$ denotes the CU depth of the $B_{i,M}$ representation. In this scheme, larger depths are skipped in the RDO process of the dependent representations, and considerable time is saved. For example, if the optimal CU depth is calculated to be 1 (*i.e.*, $32 \times 32$ pixels) for a block in the highest bitrate representation, CUs with depth 2 (*i.e.*, $16 \times 16$ pixels) and depth 3 (*i.e.*, $8 \times 8$ pixels) are skipped from the RDO process when encoding the co-located CUs in dependent representations. It means that the overall encoding time is bound by the encoding time of the highest resolution encode. Thus, this approach is not efficient in the parallel encoding environment.

**MR – Scheme 1:**

This scheme is based on the heuristic that the CU depth in the intermediate bitrate representations is highly likely to be between the CU depth of the highest and lowest bitrate representations. Thus, this scheme uses the information from the highest and the lowest bitrate representations to reduce the encoding runtime complexity of the intermediate representations.

First, the highest bitrate representation is encoded as a stand-alone encoding (*i.e.*, independent of RDO of other representations). The CU depth information obtained from this encoding is used as the upper bound of the CU depth estimation process of the lowest bitrate representation similar to [7]. The remaining intermediate quality representations are encoded with a double bound approach for CU depth search as shown in Figure 4. Hence, the possible depth level search window is limited by the depth values of the highest and lowest bitrate representations, as shown in Equation 2.

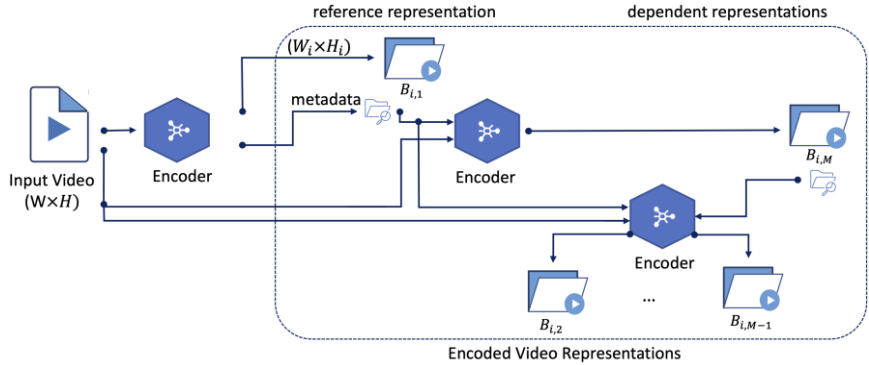

Figure 4. A double bound approach to further reduce the time complexity of intermediate representations.

$$[d_L, d_U] = \begin{cases} [0, 3], & B_{i,M} \\ [0, d_{i,M}], & B_{i,1} \\ [d_{i,1}, d_{i,M}], & B_{i,j} \forall j \in [2, M-1] \end{cases} \quad (2)$$

**MR – Scheme 2:**

In parallel encoding, the overall time-complexity is limited to the maximum time-complexity of one of the representations that are encoded in parallel. Therefore, as shown in Figure 5, it is proposed to encode the lowest bitrate first and use its information to reduce the highest time-complexities. To this end, the information of the co-located CU in the reference encoding and raw video is fed to a *Convolutional Neural Network* (CNN) (see Figure 6) to predict split decisions for CUs.

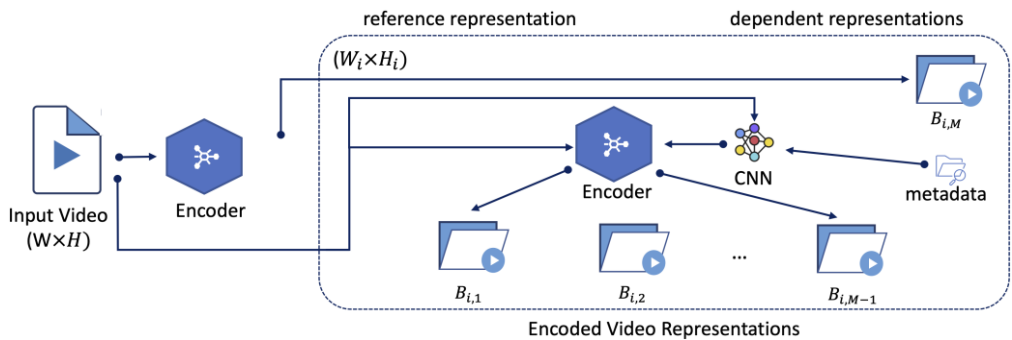

Figure 5. A multi-rate encoding scheme designed for parallel encoding scenarios.

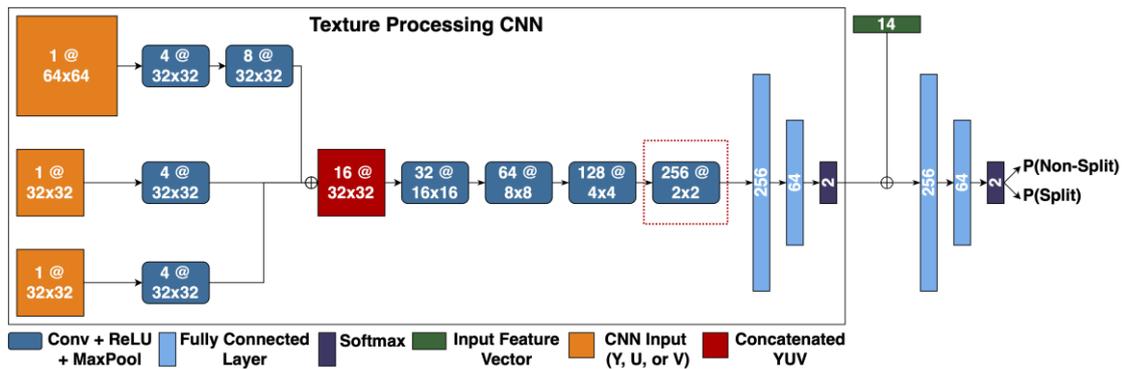

Figure 6. Network architecture used for depth 0 classifier. The numbers in the boxes are in the following format from left to right: $Channel\ count\ @\ Width \times Height$ of channel for convolution layers, *output size* for fully connected and softmax layers, and *input size* for the feature vector. Red dotted section is removed in the depth 1 classifier due to variance in the input size. Y, U, and V input sizes and intermediate channel sizes vary depending on the depth level (halved for depth 1 classifier).

### MR – Scheme 3:

In this scheme, specific proposed heuristics that can be applied coupled with CU depth search optimization-based multi-rate encoding approaches to improve the overall encoding speed without compromising the coding efficiency are introduced.

### *Prediction Mode Heuristics:*

If the CU is split to the CU size chosen in the highest bitrate representation, the prediction mode heuristics are proposed for the remaining representations as follows:

(i) It is observed that if Skip 2Nx2N mode was chosen as the best mode in the RDO process of the highest bitrate representation, Skip 2Nx2N mode is likely to be selected in the other representations. Hence, if Skip 2Nx2N mode was selected in the highest bitrate representation, RDO is evaluated for only Merge/Skip 2N×2N and Inter 2N×2N modes.

(ii) It is observed that if 2Nx2N mode was selected as the best mode in the RDO process of the highest bitrate representation, it is less likely that an asymmetrical motion prediction (AMP) mode would be selected in the other representations. Hence, If the 2Nx2N mode was chosen in the highest bitrate representation, RDO is skipped for AMP modes.

(iii) It is observed that if any inter-prediction mode was selected as the best mode in the RDO process of the highest bitrate representation, it is less likely that any intra-prediction mode would be selected in the other representations. Hence, if any inter-prediction mode was chosen in the highest bitrate representation, RDO is skipped for intra-prediction modes.

(iv) It is observed that the probability of PUs in intermediate bitrate representations choosing an intra-prediction mode is very high when the co-located PUs in the highest and lowest bitrate representations have selected an intra-prediction mode. Hence, if an intra-prediction mode was chosen for the highest and the lowest bitrate representations, RDO is evaluated for only Merge/Skip 2N×2N and intra-prediction modes in the intermediate representations.

*Motion Estimation Heuristics:*

If the CU is split to the CU size chosen in the highest bitrate representation and the PU size is also the same as that of the highest bitrate representation, the motion estimation heuristics are proposed for the remaining representations as follows:

(i) It is observed that the probability of the dependent representations choosing the same reference frame as that is the highest bitrate representation is very high. Hence, the same reference frame is selected as the highest bitrate representation, and other reference frame searches are skipped in the dependent representations.

(ii) The Motion Vector Predictor (MVP) is set to be the Motion Vector (MV) of the highest bitrate representation.

(iii) The motion search range is decreased to a smaller window size if the MV of the highest bitrate representation and the MV of the lowest bitrate representation are close. The search range is determined to be the maximum difference between the x and y coordinates of the MVs.

An efficient multi-rate encoding scheme is designed tailor-made for parallel encoding environments. In this scheme, the lowest bitrate representation is encoded first. The CU depth search in the highest bitrate representation is lower bound by the CU depth values of the lowest bitrate representation. The remaining intermediate quality representations are encoded with a double-bound approach for CU depth search and the proposed heuristics. This scheme is proposed specifically to encode multiple bitrate representations of a single resolution in parallel. The encoder analysis flowchart for this scheme is similar to Figure 5 but without using CNN and with using the above-mentioned heuristics.

**MULTI-ENCODING (ME) SCHEMES**

Schroeder et al. extend their multi-rate encoding approach in [7] to a multi-encoding approach in [16], where the highest bitrate representation of the highest resolution is encoded first and its information is reused not only to speed of other representations with the same resolution but also representations with different resolutions. Figure 7 shows the workflow for this approach. Since the overall encoding time is bound by the encoding time of the highest resolution encode, this approach is not efficient in the parallel encoding environment.

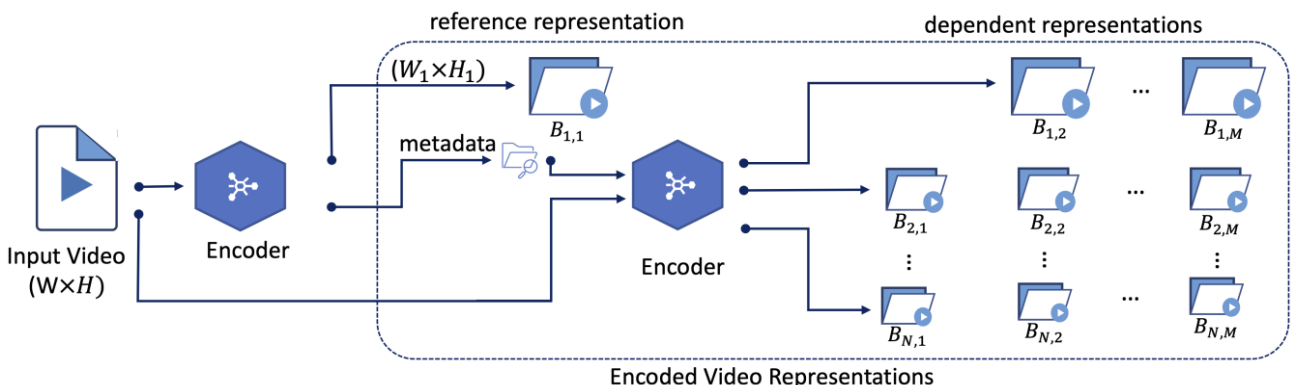

Figure 7. Encoder analysis flow diagram of [16].

In this Section, multi-encoding schemes based on machine learning (ME – Ischeme 1), and the proposed heuristics and multi-resolution approaches are introduced. ME – Scheme 2 is optimized for the highest compression efficiency, ME – Scheme 3 is optimized for the best compression efficiency encoding time savings trade-off, and ME – Scheme 4 is optimized for the highest encoding time savings. The optimizations are skipped for intra-coded CTUs in the following schemes since the contribution to the encoder speedup from those optimizations is insignificant.

**ME – Scheme 1:**

A machine learning-based approach is proposed for fast multi-encoding for HAS. As shown in Figure 8, first, the highest quality representation from the lowest resolution is encoded and it is chosen as the reference representation. Encoding information (*e.g.*, RD cost, intra directions, motion vectors, prediction modes, etc.) from the reference representation in addition to luma (*i.e.,* Y), chroma (*i.e.,* U, and V) values from the target resolution for a given CTU are passed to the CNN to predict the split decision. A CU split decision dataset is constructed to train separate CNNs for each target QP and resolution combination. The trained CNN is then applied during the encoding of dependent representations.

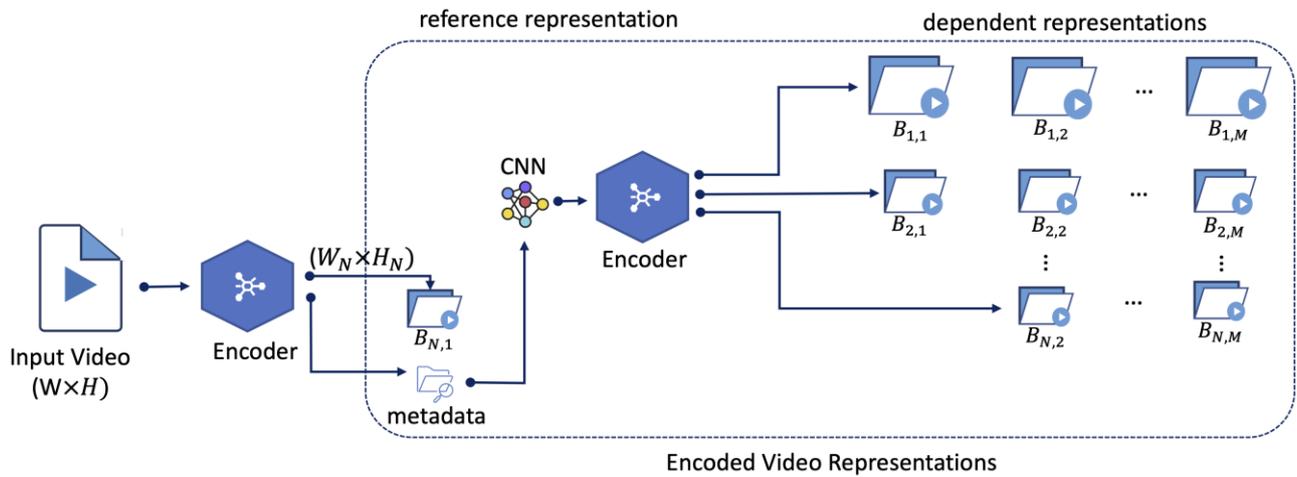

Figure 8. Encoder analysis flow diagram of proposed learning based multi-encoding scheme.

**ME – Scheme 2:**

ME – Scheme 2 is used for encoder analysis sharing across representations within a resolution. As the multi-resolution approach, the CU depth information from the highest bitrate representation of the $(i-1)^{th}$ resolution, $d_{i,M}$ is shared to the highest bitrate representation of the $(i)^{th}$ resolution. The lower-bound for CU depth estimation in the highest bitrate representation of the $(i)^{th}$ resolution representation, *i.e.*, $d_L$ is determined as:

$$d_L = \begin{cases} d_{i-1,M} - 1 & if\, d_{i-1,M} \geq 1 \\ 0, & \text{otherwise} \end{cases}$$

The encoder analysis flow diagram of this scheme is shown in Figure 9.

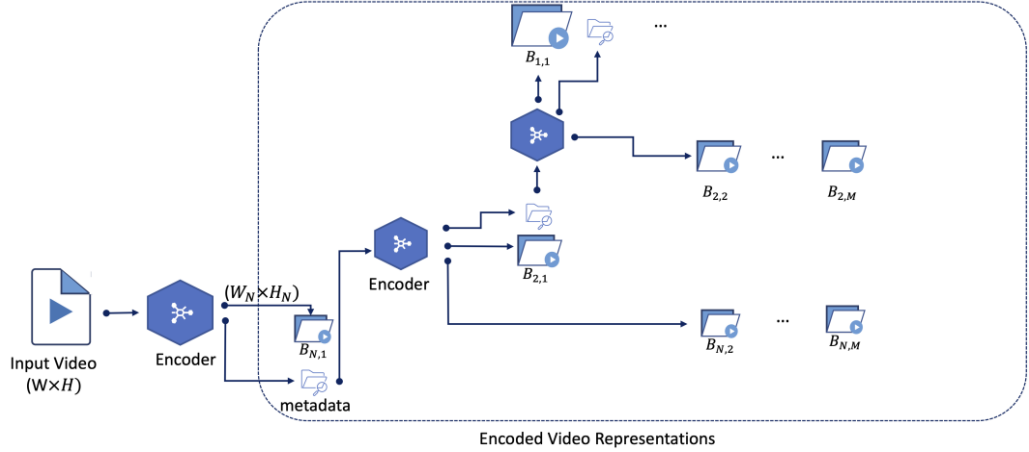

Figure 9. Encoder analysis flow diagram of the proposed multi-encoding scheme (ME – Scheme 2).

**ME – Scheme 3:**

In ME – Scheme 3, the CU depth information from the lowest bitrate representation of the $(i-1)^{th}$ resolution, $d_{i,1}$ is shared to the highest and lowest bitrate representations of the $(i)^{th}$ resolution. The lower-bound for CU depth estimation in the highest and lowest bitrate representations of the $(i)^{th}$ resolution, i.e., $d_L$ is determined as:

$$d_L = \begin{cases} d_{i-1,1} - 1, & d_{i-1,1} \geq 1 \\ 0, & \text{otherwise} \end{cases} \quad (3)$$

The encoding analysis flow diagram of this scheme is similar to Figure 9 with the difference of replacing reference representation in each resolution with the lowest bitrate representation.

**ME – Scheme 4:**

ME – Scheme 4 aims to maximize the encoding time savings in parallel encoding by decreasing the encoding times of the highest bitrate representations of resolution layers. dL is first computed for the highest bitrate representation of the $(i)^{th}$ resolution. The PU sizes, MVs are scaled by the resolution increase factor $L_i$ defined as:

$$L_i = \frac{W_i \times H_i}{W_{i-1} \times H_{i-1}} \quad (4)$$

where Wi and Hi represent the width and height of video frame at the $(i)^{th}$ resolution. The CU depth, scaled PU size, and mode decisions are reused in the highest bitrate representation of the $(i)^{th}$ resolution. dL is re-evaluated by using the RDO against the cost of splitting the CU (i.e., $d_L$ + 1) by computing the optimal PU modes. At the same time, the scaled MVs are reused as motion vector predictors for the co-located PUs in the $(i)^{th}$ resolution.

## EXPERIMENTAL RESULTS

This section first introduces the test methodology used in this paper and then presents the experimental results.

**Test Methodology:**

All schemes presented in this paper are implemented using x265 v3.5 [17] with the *veryslow* preset and the Video Buffering Verifier (VBV) rate control mode. Psycho-visual optimizations and adaptive quantization are not used in the evaluation. For encoder analysis sharing in the considered schemes, per-segment encoding analysis metadata (file) is generated by the reference representations along with the HEVC bitstream and shared to the dependent representations. All experiments are run on a dual-processor server with Intel Xeon Gold 5218R (80 cores, frequency at 2.10 GHz) which utilizes multi-threading optimizations [18] (*i.e.,* Wavefront Parallel Processing (WPP) and frame-threading) and x86 SIMD optimizations [19] of x265. ABR ladder encoding of four test sequences was run in parallel. In this configuration, the full multi-threading capabilities of x265 are utilized. Video sequences from JVET [20], MCML [21], and SJTU [22] datasets are used, representing various types of contents. As mentioned in Table 1, twelve representations are considered in the ABR ladder: three resolutions (N = 3) with four bitrates for each resolution (M = 4). The bitrates are chosen in compliance to HTTP Live Streaming (HLS) specification [23].

Table 1: Bitrate Ladder.

| Bitrate | 540p (i=1) | 1080p (i=2) | 2160p (i=3) |
|---|---|---|---|
| $B_{i,1}$ | 0.5 Mbps | 3.0 Mbps | 11.6 Mbps |
| $B_{i,2}$ | 1.0 Mbps | 4.5 Mbps | 16.8 Mbps |
| $B_{i,3}$ | 1.5 Mbps | 5.8 Mbps | 20.0 Mbps |
| $B_{i,4}$ | 2.0 Mbps | 7.0 Mbps | 25.0 Mbps |

When encoding multiple representations serially, *i.e.,* one after the other, the overall encoding time-complexity, $\tau_S$ is the sum of the encoding times of all representations as shown in Equation 5.

$$\tau_S = \sum_{i=1}^{N} \sum_{j=1}^{M} (\tau_{B_{i,j}}) \qquad (5)$$

where N denotes the number of resolutions, and M denotes the number of bitrate representations in the $(i)^{th}$ resolution. $\tau_{B_{i,j}}$ represents the time taken for encoding the representations, $B_{i,j} \forall i \in [1, N]$ and $j \in [1, M]$. When encoding multiple representations in parallel, the overall encoding time-complexity is bounded by the encoding time-complexity of the highest bitrate representation ($B_{N,M}$) [24]. Therefore, although utilizing the highest bitrate representation as the reference will reduce encoding time-complexity for dependent representations, the overall encoding time will remain unchanged if all representations are encoded in parallel. The overall encoding time depends on the encoding time of the highest bitrate representation. This is represented in Equation 6, where $\tau_P$ denotes the total time taken for parallel encoding.

$$\tau_P = max_{i,j}(\tau_{B_{i,j}}); j \in [1, M], i \in [1, N] \qquad (6)$$

The resulting encoding time, quality in terms of PSNR and VMAF [25], and the achieved bitrate are compared for each test sequence. Since, it is assumed that representations are displayed on the highest resolution, *i.e.,* 2160p, all representations are scaled (bi-cubic) to 2160p to calculate VMAF and PSNR [2,3]. In the experimental results, $\Delta T_S$ and $\Delta T_P$ represent the cumulative encoding time saving for all bitrate representations compared to the stand-alone encoding in serial and parallel encoding scenarios, respectively; Bjøntegaard delta rates $BDR_P$ and $BDR_V$ refer to the average increase in bitrate of the representations to that of the stand-alone encoding to maintain the same PSNR and VMAF respectively. A positive

BDR indicates a drop in coding efficiency of the proposed method compared to the stand-alone encoding, while a negative BDR represents a coding gain. Table 2 provides the summary analysis of the encoding time saving ($\Delta T$) and BDR, respectively. For multi-rate encoding schemes only 2160p resolution has been considered.

Table 2: Results of the multi-rate and multi-encoding schemes.

| Scheme | $\Delta T_S$ | $\Delta T_P$ | $BDR_P$ | $BDR_V$ |
|---|---|---|---|---|
| MR - Scheme 1 | 25.92% | 0% | 1.86% | 2.22% |
| MR - Scheme 2 | 38.12% | 29.89% | 1.47% | 1.84% |
| MR - Scheme 3 | 37.89% | 20.74% | 4.50% | 4.42% |
| ME - Scheme 1 | 46.27% | 18.14% | 1.96% | 2.16% |
| ME - Scheme 2 | 34.71% | 22.03% | 2.37% | 2.38% |
| ME - Scheme 3 | 45.27% | 20.72% | 3.32% | 3.17% |
| ME - Scheme 4 | 68.76% | 76.82% | 4.51% | 4.48% |

**CONCLUSIONS**

This paper described the challenges within the HAS encoding framework of the widely used HEVC codec, and how the encoder analysis sharing within the encoding representations reduces the encoding time without significant drop in the overall compression efficiency. This paper described various multi-rate and multi-encoding schemes used with HM and x265 HEVC open-source encoders, which enables efficient adaptive streaming of video content. The fastest scheme proposed in this paper also includes efficient sharing of inter-frame motion structures and the ability to refine shared data across encoder instances.

**ACKNOWLEDGEMENTS**

The financial support of the Austrian Federal Ministry for Digital and Economic Affairs, the National Foundation for Research, Technology, and Development, and the Christian Doppler Research Association is gratefully acknowledged. Christian Doppler Laboratory ATHENA: https://athena.itec.aau.at/